\documentclass[12pt, a4paper]{article}

\usepackage{ifpdf}
\ifpdf
\usepackage{cmap} % возможность поиска по pdf
\usepackage[pdftex]{color, graphicx} % использование цвета и вставка графики
\usepackage{epstopdf} % автоматическая конвертация картинок eps в pdf
\usepackage[pdftex, colorlinks=false, pdfstartview=FitH, hypertexnames=false, unicode, bookmarks=false]{hyperref} % автоматическое создание гиперссылок
\else
\usepackage[dvips]{color, graphicx} % использование цвета и вставка графики
\usepackage[hypertex, colorlinks=false, unicode]{hyperref} % автоматическое создание гиперссылок
\fi

\graphicspath{{Pics/}{Pics/More/}} % дополнительные пути для поиска графических файлов

\usepackage[cp1251]{inputenc} % кодировка текущего документа
\usepackage[english]{babel} % загружаемые языки, последний имеет приоритет

\usepackage{amssymb, amsmath, amsxtra}

\usepackage{indentfirst}

\begin{document}

\date{}

\title{
Direct URCA Processes in Supernovae and \\ 
Accretion Disks with Arbitrary Magnetic Field
}

\author{
Igor Ognev{\footnote{e-mail: ognev@uniyar.ac.ru}}
\\[2mm]
{\it P.\,G. Demidov Yaroslavl State University}
\\%	[2mm]
{\it Sovetskaya 14, 150003 Yaroslavl, Russia}
}

\maketitle

\vspace*{-5mm}
\begin{abstract}{\footnotesize
An effect of a magnetic field of an arbitrary strength on the beta-decay and reactions 
related with it by the crossing symmetry (the beta-processes) in supernovae and accretion 
disks around black holes is analyzed. Rates of the beta-processes and the energy and 
momentum transfered through them to an optically transparent matter are calculated. It is 
shown that the macroscopic momentum transferred to the medium increases linearly with the 
magnetic field strength and can substantially affect the dynamics of supernovae and accretion 
disks especially when a matter inside is degenerate. It is also demonstrated that the rates 
of the beta-processes and the energy deposition in these reactions for the magnetic field 
strength $B \lesssim 10^{15}$~G, which is assumed to be typical in supernovae and accretion 
disks, are lower than in the absence of the field. This suppression is more pronounced 
for reactions with neutrinos.
}\end{abstract}

\section*{Introduction}
\label{intro}

Neutrinos can strongly affect the processes occurring in many astrophysical objects, especially 
explosive processes accompanied by the release of a large amount of energy and heating of the 
surrounding medium. Such processes include the core collapse supernova which is the final stage 
in an evolution of massive stars as well as a merger of close binary systems. Despite the fact 
that both these phenomena are fundamentally different, their common features are the powerful 
neutrino radiation and possible connection with cosmological gamma-ray bursts~\cite{Kumar:2014upa} 
which remain an unsolved problem in theoretical astrophysics for many years. In a supernova 
explosion, a source of the neutrino emission is the proto-neutron star that forms as the result 
of the core collapse of a massive star at the final stage of its evolution. It should be noted 
that neutrinos are the main engine leading to a successive explosion of supernovae. A merger 
of stars in a close binary system leads to the formation of a central object (as a rule, it 
is a black hole or, very rarely, a massive neutron star) and a hot accretion disk around it. 
In this case, the disk is a source of powerful fluxes of neutrinos. Note that an analogous system 
consisting of a central black hole and accretion disk can be formed during a failed supernova 
explosion called the collapsar model~\cite{MacFadyen:1998vz}. In this model, a black hole 
is appeared as the result of the core collapse of a massive star, rather than a proto-neutron 
star. The absence of a powerful neutrino radiation from the central part leads to the accretion 
of the remaining matter to the black hole and formation of a hot disk around the hole. General 
interest in such systems is connected with the fact that a large amount of energetic neutrino 
is emitted from the hot disk and produces a relativistic electron-positron plasma which could 
be a source of gamma-ray bursts~\cite{Kumar:2014upa}. However, the merger of compact objects 
(neutron stars or black holes) apparently leads to the formation of short-duration gamma-ray 
bursts~\cite{Rosswog:2015nja}. Long-duration gamma-ray bursts are usually associated with the 
supernova explosions with a large energy release and probably with collapsars~\cite{Kumar:2014upa}. 
It should be noted that the merger of a compact object with the massive Wolf-Rayet star can also 
be the source of the long-duration gamma-ray bursts~\cite{Barkov:2009zq}.

The dense and hot medium that forms after a supernova explosion, as well as the merger of stars 
in a close binary system, is not transparent to photons, and emission of neutrinos is the main 
channel for its cooling. Due to high temperatures, the medium predominantly consists of free 
nucleons, photons, and relativistic electrons and positrons being in the local thermodynamic 
equilibrium. Depending on the specific conditions, neutrinos in such a medium can either be  
in equilibrium, like in supernova cores, or be freely emitted, like in accretion disks or 
envelopes of supernovae. The main neutrino reactions in such media are the beta-processes 
(called often the direct URCA-processes)~\cite{Bruenn:1985en}:
\begin{gather}
\label{proc:pe}
p + e^- \rightleftarrows n + \nu_e , 
\\
\label{proc:ne}
n + e^+ \rightleftarrows p + \bar\nu_e , 
\\
\label{proc:n}
n \rightleftarrows p + e^- + \bar\nu_e ,
\end{gather}
which are important primarily as the main channel of the energy exchange between 
the medium and neutrinos in the case of supernovae and the energy source of gamma-ray 
bursts in the merger of stars in the close binary system. An important feature in such 
astrophysical cataclysms is that a strong magnetic field can be generated. Simulations 
show that the magnetic field strength in supernovae~\cite{Moiseenko:2015,Sawai:2015tsa} 
and in accretion disks~\cite{Barkov:2009zq,Zalamea:2010ax} can reach the values 
$B \sim 10^{15}$~G. Such a strong magnetic field not only significantly affects 
the dynamics of supernovae and accretion disks but also modifies the interaction 
of neutrinos with their matter 
(see, for example,~\cite{Gvozdev:2002ta,Gvozdev:2002nu,Gvozdev:2005hz}).

In this paper, the modifications of the beta-processes taking place in supernovae 
and accretion disks, by the magnetic field of an arbitrary strength are discussed.  
The results of calculations of the neutrino- and antineutrino-production rates 
and the energy-momentum deposition to optically transparent medium are presented. 
A dependence of these quantities on the magnetic field strength and parameters 
of the medium is demonstrated. Other details of the processes presented here  
can be found in the recent paper~\cite{Ognev:2016wlq}.

In this paper we use the system of units in which $c = \hbar = k = 1$, 
where $c$~is the velocity of light, $\hbar$~is the Planck constant,
and $k$~is the Boltzmann constant.

\section{General formalism}
\label{sec-1}

The macroscopic quantities, being the most interesting for astrophysical applications, 
are the reaction rate (the number of processes occurring in a unit volume per unit time):
\begin{equation}
\label{eq:G}
\varGamma 
=  
\frac{1}{V} \int 
\sum_{i, \; f} \frac{\left | S_{if} \right |^2}{{\cal T}}
\prod_{i, \; f} f_i \left ( 1 - f_f \right ) dn_i \, dn_f ,
\end{equation}
and the energy and momentum transferred from neutrinos 
to a unit volume of the medium per unit time:
\begin{equation}
\label{eq:Q}
Q^\mu 
= \left ( Q, \vec{\mathcal{F}} \right ) =
\pm \frac{1}{V} \int 
\sum_{i, \; f} q^\mu \, \frac{\left | S_{if} \right |^2}{{\cal T}}
\prod_{i, \; f} f_i \left ( 1 - f_f \right ) dn_i \, dn_f ,
\end{equation}
where~$V$ is the normalization volume,
$\left | S_{if} \right |^2 \!\!/ {\cal T}$ is the transition probability 
from the initial~($i$) state to final~($f$) one per unit time, 
$q^\mu = (\omega, \vec q)$ is the neutrino four-momentum, and the sign~"$\pm$" 
in~(\ref{eq:Q}) specifies either absorption or emission neutrino process 
is considered. The summation and integration is carried out over the phase 
space~$dn_{i,f}$ of all particles taking into account their distribution 
functions~$f_{i,f}$ (it is implicitly assumed that all final particles 
are fermions). It is convenient to separate the integration over the 
neutrino momentum from all other. One can define the quantity called 
the emissivity~\cite{Ognev:2016wlq}: 
\begin{equation}
\label{eq:K}
\mathcal{K}(q)
=
\sum_{i, \, f \neq \nu} \int 
\frac{\left | S_{if} \right |^2}{{\cal T}} 
\prod_{i, \, f \neq \nu} f_i \left ( 1 - f_f \right ) dn_i \, dn_f ,
\end{equation}
where the reaction with neutrino in the final state is assumed. 
It completely determines the property of the matter to emit or absorb 
the neutrino with the four-momentum~$q$ in the transition $i \leftrightarrow f$.

Under conditions of supernovae and accretion disks, the total emissivity 
for neutrino and antineutrino in beta-processes can be presented in the 
simple analytical form~\cite{Ognev:2016wlq}:
\begin{equation}
\label{eq:K_nu}
\begin{gathered}
\mathcal{K}_{\nu} (q) 
=
K_0 \, \varPhi (b, x_\nu, \beta_\nu) 
\, \frac{e^{(\delta\chi + \Delta) / t - a}}{e^{x_\nu / t - a} + 1}  ,
\\
\mathcal{K}_{\bar\nu} (q) 
=
K_0 \, \varPhi (b, x_{\bar\nu}, \beta_{\bar\nu}) 
\, \frac{1}
     {e^{x_{\bar\nu} / t + a} + 1}  ,
\\
K_0
=
\frac{\left ( g_v^2 + 3 g_a^2 \right ) \cos^2\!\theta_c}{2 \pi} \, 
G_F^2 \, N_n \, m_e^2  ,
\\
\begin{aligned}
\varPhi &(x, b, \beta)
=
2 \, x \, \sqrt{x^2 - 1} \ \theta \big ( x - \sqrt{1 + 2b} \big ) \, +
\\
& + \frac{b \, x}{\sqrt{x^2 - 1}} \, \Big [
\theta \big ( x - 1 \big )  - 
\theta \big ( x - \sqrt{1 + 2b} \big ) 
\Big ]
%- \frac{g_a^2 - g_v^2}{3 g_a^2 + g_v^2} \, 
- g_{va} \, \frac{b \, x \, \cos\beta}{\sqrt{x^2 - 1}} 
\ \theta \big ( x - 1 \big )  ,
\end{aligned}
\end{gathered}
\end{equation}
where $m_n$, $m_p$, $m_e$ and $\mu_n$, $\mu_p$, $\mu_e$ are the masses 
and local chemical potentials of neutrons, protons and electrons,
$\Delta = (m_n - m_p) / m_e$, $\delta\chi = (\mu_e + \mu_p - \mu_n) / m_e$,
$T$ is the local temperature of the medium, 
$t = T / m_e$, $a = \mu_e / T$, 
$N_n$ is the neutrons number density,
$x_{\nu, \bar{\nu}} = \omega_{\nu, \bar{\nu}} / m_e \pm \Delta$ 
are the scaled neutrino and antineutrino energy,
$b = eB / m_e^2 = B / B_0$ is the magnetic field strength measured 
in units of the Schwinger value $B_0 \approx 4.41 \times 10^{13}$~G,  
$e > 0$ is the elementary charge, 
$\beta_{\nu, \bar{\nu}}$ is the angle between the (anti)neutrino 
momentum and magnetic field direction.
Here, the following constants are used~\cite{Olive:2016xmw}:
$g_v \!\approx\! 1$, $g_a \!\approx\! -1.27$ 
are the vector and axial constants of the nucleon charged current,
$g_{va} \!= (g_a^2 - g_v^2) / (3 g_a^2 + g_v^2) \!\approx\! 0.11$,
$\theta_c$ is the Cabibbo angle, $\cos\theta_c \!\approx 0.97$,
$G_F \approx 1.17 \times 10^{-11}$~MeV$^{-2}$ is the Fermi constant.
The dependence on the magnetic-field strength is entering 
in~$\mathcal{K}_{\nu} (q)$ and~$\mathcal{K}_{\bar\nu} (q)$ through 
the function $\varPhi (x, b, \beta)$ only. The first term in this 
function coincides with the one appeared in the absence of the field 
and describes the magnetized matter ability to emit and absorb 
neutrinos with high energies ($x \geqslant\! \sqrt{1 + 2b}$). 
The second term in $\varPhi (x, b, \beta)$ dominates in the limit 
of the strong magnetic field and responsible for the low-energy 
neutrino spectrum ($x \leqslant\! \sqrt{1 + 2b}$). Note that this
term is the result of exact calculations~\cite{Ognev:2016wlq} 
of neutrino and antineutrino emissivities for the beta-processes 
under the supernova and accretion disk conductions. The last term 
in $\varPhi (x, b, \beta)$ is linear in $\cos\beta$ and determines 
the asymmetric part of the emissivity. It existence leads to an 
uncompensated momentum transfered to the medium along the magnetic 
field direction.   %   at the expense of the beta-processes.

\section{Medium transparent for neutrino}
\label{sec-2}

The case of a medium transparent for neutrinos is the most simple 
for calculations. However, it is quite applicable in astrophysics. 
In particular, the matter of accretion disks and an external part 
of supernova envelopes is transparent for neutrinos and antineutrinos 
and their distribution functions can be well approached by unit, 
$f_{\nu, \bar{\nu}} \ll 1$, in~(\ref{eq:G}) and~(\ref{eq:Q}).
In this case, the neutrino- and antineutrino-production number and 
the energy emitted from the unit volume of the medium per unit time 
in the beta-processes can be written as follows: 
\begin{equation}
\begin{gathered}
\label{GQ}
\varGamma_{\nu, \bar\nu}
=
\frac{1}{(2 \pi)^3} \int \mathcal{K}_{\nu, \bar\nu} (q) \, d^3 q ,
\\
Q_{\nu, \bar\nu}
=
\frac{1}{(2 \pi)^3} \int \omega \,
\mathcal{K}_{\nu, \bar\nu} (q) \, d^3 q . 
\end{gathered}
\end{equation}
As noted above, the existence of the asymmetric part in the emissivity 
(the third term in $\varPhi (x, b, \beta)$ in Eq.~(\ref{eq:K_nu}))
leads to a macroscopic momentum transferred to the medium. In the case 
of the transparent for (anti)neutrino matter only the component along 
the magnetic field direction is non zero:
\begin{equation}
\mathcal{F}_{\nu, \bar\nu \, \|}
=
- \frac{1}{(2 \pi)^3} \int \omega \, \cos\beta \,
\mathcal{K}_{\nu, \bar\nu} (q) \, d^3 q . 
\end{equation}
This quantity becomes quite simple in the case of the ultrarelativistic 
electron-positron plasma which exists in supernovae and accretion disks. 
In this limit, the force density along the magnetic field arising from 
neutrino and antineutrino emission has the form~\cite{Ognev:2016wlq}: 
\begin{equation}
\label{eq:F-nu-long}
\begin{gathered}
\mathcal{F}_{\nu \, \|}
=
\mathcal{F}_0 \, I_3(a) \, Y \, / \, (1+Y) ,
\quad 
\mathcal{F}_{\bar\nu \, \|}
=
\mathcal{F}_0 \, I_3(-a) \, / \, (1+Y) ,
\\
\mathcal{F}_0 = 
\frac{\left( g_a^2 - g_v^2 \right) \cos^2\!\theta_c}{12 \pi^3} 
\, G_F^2 \, eB \, T^4 \frac{\rho}{m_N}  ,
\quad 
I_3(a)
=
\int\limits_0^\infty \frac{y^3 \, dy}{e^{y-a} + 1}  , 
\end{gathered}
\end{equation}
where $Y \!=\! N_p / N_n$ is the proton-to-neutron number density ratio,
$\rho$ is the density of the matter, 
$m_N \!\approx\! 940$~MeV is the nucleon mass~\cite{Olive:2016xmw}.
Here, $\mathcal{F}_0 \approx 1.85 \times 10^{14} \, b \, t^4 \rho_{12} \ \text{dyn} / \text{cm}^{3}$ 
gives the typical value   %   order of magnitude 
for the force density in the nondegenerate electron-positron plasma 
with $\rho_{12} = \rho / 10^{12}\ \text{g} / \text{cm}^{3}$.
In the limit of large values of~$a$, the integral $I_3 (a)$ 
in~(\ref{eq:F-nu-long}) can be approximated by the following 
expressions~\cite{Ognev:2016wlq}:    %  ($a \geqslant 0$):
\begin{equation}
\label{eq:I3-large-a}
I_3 (a) 
\approx 
\frac{a^4}{4} + \frac{\pi^2}{2} \, a^2 + 3.2 \, a + \frac{7 \pi^4}{120} ,
\quad\quad 
I_3 (-a) 
\approx 
6 \, e^{- a}  .
\end{equation}
So, for large values of the electron degeneracy parameter $a = \mu_e/T$ 
the momentum transferred to the medium increases by the factor $\sim a^4$ 
for the processes with the neutrino participation and exponentially 
suppressed for reactions with antineutrino.

\begin{figure}[htbp]
\vspace{2mm}
\begin{minipage}[t]{0.49\textwidth}
\centering{
\includegraphics[width=0.95\textwidth]{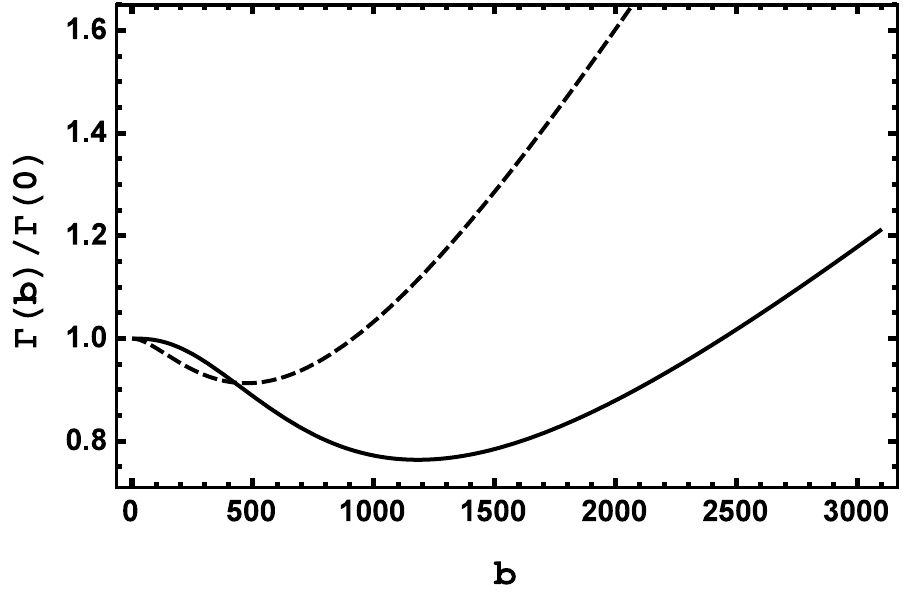} 
}
\end{minipage}
\hfill
\begin{minipage}[t]{0.49\textwidth}
\centering{
\includegraphics[width=0.95\textwidth]{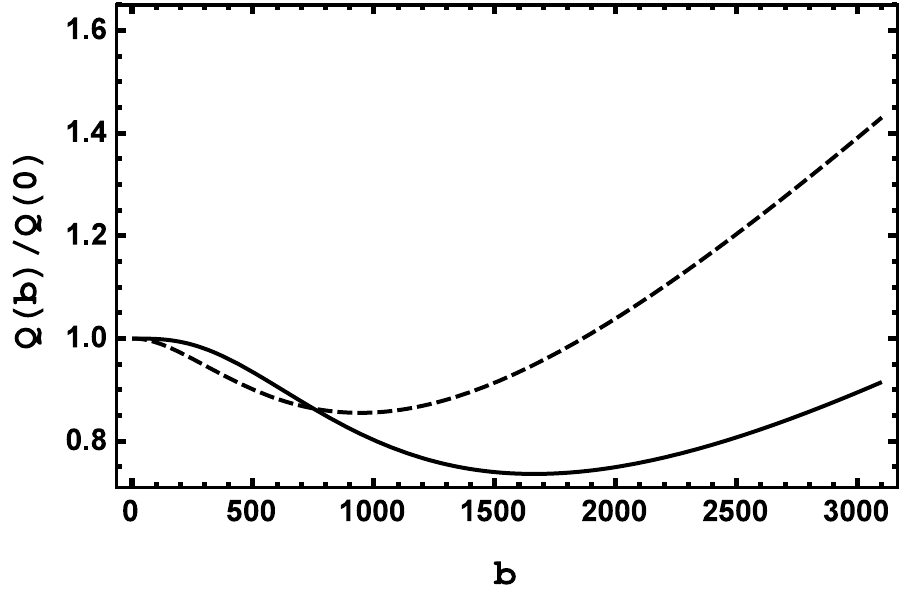}  
}
\end{minipage}
\caption{
The beta-process rates (the left panel) and energy emitted through (anti)neutrino 
in these processes (the right panel) are plotted for the dimensionless temperature 
$t \!=\! 8$ and electron degeneracy parameters $a \!=\! 3$ as functions of the 
magnetic field $b \!=\! B / B_0$. The processes with the neutrino and antineutrino 
participation are shown by the solid and dashed curves, respectively.  
} 
\label{fig:GQ}
\end{figure}

In the case of an ultrarelativistic electron-positron plasma, the neutrino- 
and antineutrino-production rates and the energy emitted in these processes 
depend not on the magnetic field strength itself but on dimensionless 
parameter $y_b =\! \sqrt{2b} / t =\! \sqrt{2eB} / T$ which contains 
the plasma temperature~\cite{Ognev:2016wlq}: 
\begin{equation}
\label{eq:G-Q-ultra}
\begin{gathered}
\varGamma_{\nu} 
= 
\varGamma_0 \, J_2(y_b, a) \, Y \, / \, (1+Y) ,
\quad 
Q_{\nu} 
= 
Q_0 \, J_3(y_b, a) \, Y \, / \, (1+Y) ,
\\
\varGamma_{\bar\nu} 
= 
\varGamma_0 \, J_2(y_b, -a) \, / \, (1+Y) ,
\quad 
Q_{\bar\nu} 
= 
Q_0 \, J_3(y_b, -a) \, / \, (1+Y) ,
\\
\varGamma_0 = 
\frac{\left( 3 g_a^2 + g_v^2 \right) \cos^2\!\theta_c}{8 \pi^3} 
\, G_F^2 \, T^5 \frac{\rho}{m_N} ,
\quad
Q_0 = T \, \varGamma_0 ,
\\
J_n(y_b, a)
=
y_b^2 \! \int\limits_{0}^{y_b} \frac{y^n dy}{\exp(y-a) + 1} 
+ 4 \! \int\limits_{y_b}^{\infty} \frac{y^{n+2} dy}{\exp(y-a) + 1} .
\end{gathered}
\end{equation}
Here, the values  
$\varGamma_0 \approx 9.64 \times 10^{31} \, t^5 \, \rho_{12} \ \text{cm}^{-3} \, \text{s}^{-1}$
and
$Q_0 \approx 7.88 \times 10^{25} \, t^6 \, \rho_{12} \ \text{erg} \, \text{cm}^{-3} \, \text{s}^{-1}$ 
give the typical values   %   order of magnitude 
for the neutrino- and antineutrino-production rates and the energy flux emitted 
in the beta-processes under conditions of the nondegenerate electron-positron plasma.
Obviously, the function $J_n(y_b, a)$ introduced in~(\ref{eq:G-Q-ultra}) 
is quadratic in~$y_b$ in the limit of the strong magnetic field ($y_b \!\gg\! a$). 
So, $\varGamma_{\nu, \bar\nu}$ and $Q_{\nu, \bar\nu}$ must increase linearly 
with the growth of the magnetic field in this limit. However, in the case of 
a weak magnetic field, the rate of increase in the first term in $J_n (y_b, a)$ 
is lower than the rate of the decrease in the second term. Therefore, with the 
magnetic field increased, the beta-process rates and the energy emitted in these 
processes first decrease, come through their minimum values, and only then their     
%  asymptotically 
linear increase begin as shown in Fig.~\ref{fig:GQ} for the scaled reaction rates,  
$\varGamma_{\nu, \bar\nu} (b) / \varGamma_{\nu, \bar\nu} (0)$,  
and emitted energy, $Q_{\nu, \bar\nu} (b) / Q_{\nu, \bar\nu} (0)$. 
It can be seen that the minimum for the antineutrino-production rate and energy 
emitted in such reactions is not as deep as in the processes with neutrinos and 
is reached at lower values of the magnetic field strength.

Another important fact following from Fig.~\ref{fig:GQ} should also be mentioned. 
A transition to an asymptotic linear increase for the beta-processes with 
antineutrinos occurs in lower magnetic fields in comparison to the reactions 
with neutrino. Nevertheless, the transition to asymptotic behavior even for these 
processes corresponds to the magnetic field with the strength $B \!\gtrsim\! 10^{16}$~G 
which exceeds substantially the values typical for supernovae and accretion disks. 
Consequently, the increase in the magnetic field in these objects leads not to the 
intensification but conversely to the beta-process suppression which is manifested 
more strongly for the reactions with neutrinos. In addition, the suppression of the 
energy emitted is stronger than for the rates of the beta-processes.

\begin{figure}[htbp]
\vspace{2mm}
\begin{minipage}[t]{0.49\textwidth}
\centering{
\includegraphics[width=0.95\textwidth]
{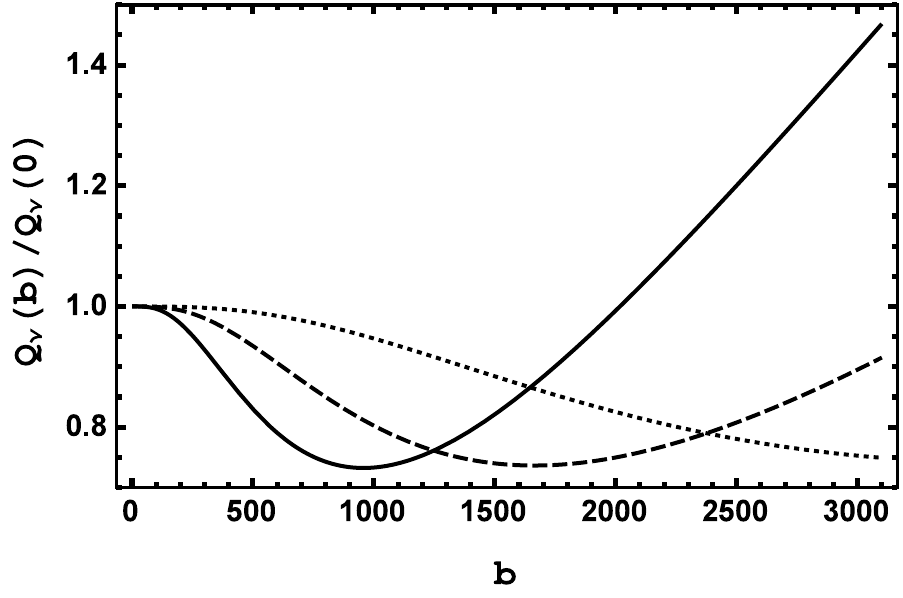} 
}
\end{minipage}
\hfill
\begin{minipage}[t]{0.49\textwidth}
\centering{
\includegraphics[width=0.95\textwidth]
{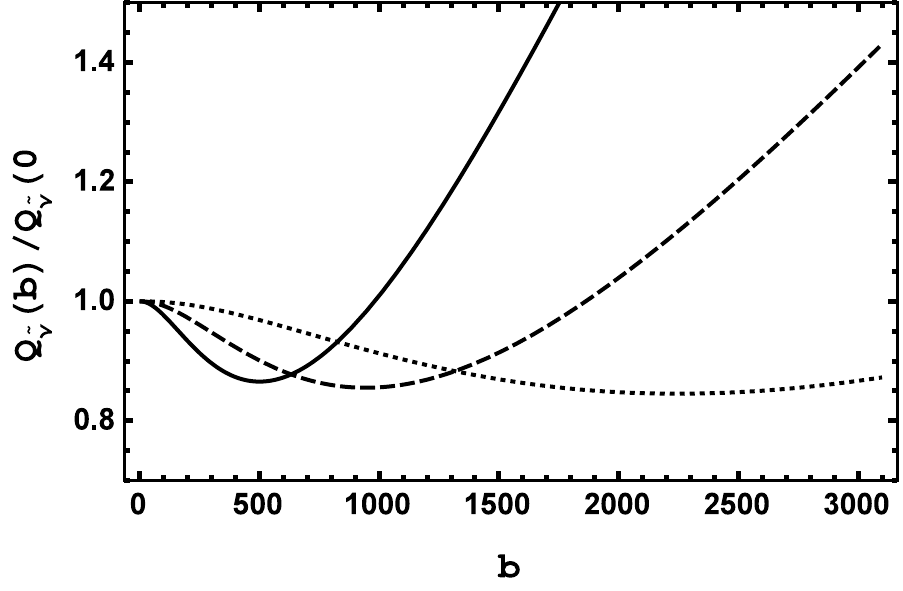} 
}
\end{minipage}
\caption{
The relative energy emitted in neutrinos (the left panel) and 
antineutrinos (the right panel), corresponding to the electron 
degeneracy parameter $a \!=\! 3$ and scaled temperatures 
$t \!=\! 6$ (solid curves), $t \!=\! 8$ (dashed curves), and 
$t \!=\! 12$ (dotted curves) as functions of the magnetic field 
strength $b \!=\! B / B_0$.
}
\label{fig:QT} 
\end{figure}

Let us consider separately the effect of other parameters on the rates of the 
beta-processes and energy emitted in them. Numerical calculations show~\cite{Ognev:2016wlq} 
that the dependence of these quantities on the temperature of the medium 
is similar for both reactions which is demonstrated in Fig.~\ref{fig:QT} 
for the relative energy $Q_{\nu, \bar\nu} (b) / Q_{\nu, \bar\nu} (0)$ 
emitted in neutrino and antineutrino.
It can be seen that with the temperature increased, the position of the minimum 
of the relative energy emitted is shifted towards stronger magnetic fields. Its 
depth for each reaction separately remains unchanged but the depth of the minimum 
for the beta-processes with the neutrino participation is larger. It should be 
noted that the dependence of the reaction rates on the temperature of the medium 
is analogous to that shown in Fig.~\ref{fig:QT} but the depth of the relative 
minimum is slightly smaller for these quantities.

A different situation is observed in the dependence of the rates of the beta-processes 
and the energy emitted in them on the electron degeneracy parameter~$a$. 
\begin{figure}[htbp]
\vspace{2mm}
\begin{minipage}[t]{0.49\textwidth}
\centering{
\includegraphics[width=0.95\textwidth]
{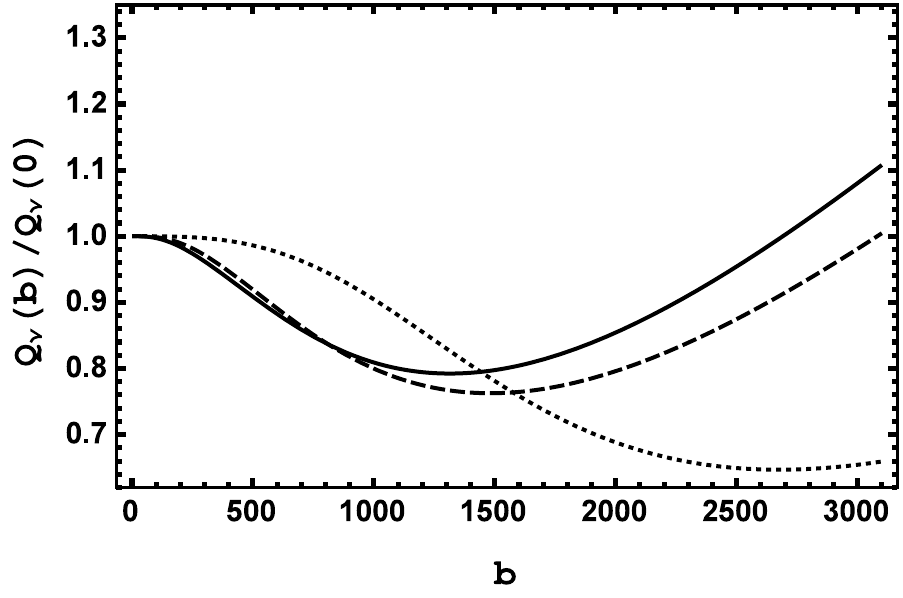} 
}
\end{minipage}
\hfill
\begin{minipage}[t]{0.49\textwidth}
\centering{
\includegraphics[width=0.95\textwidth]
{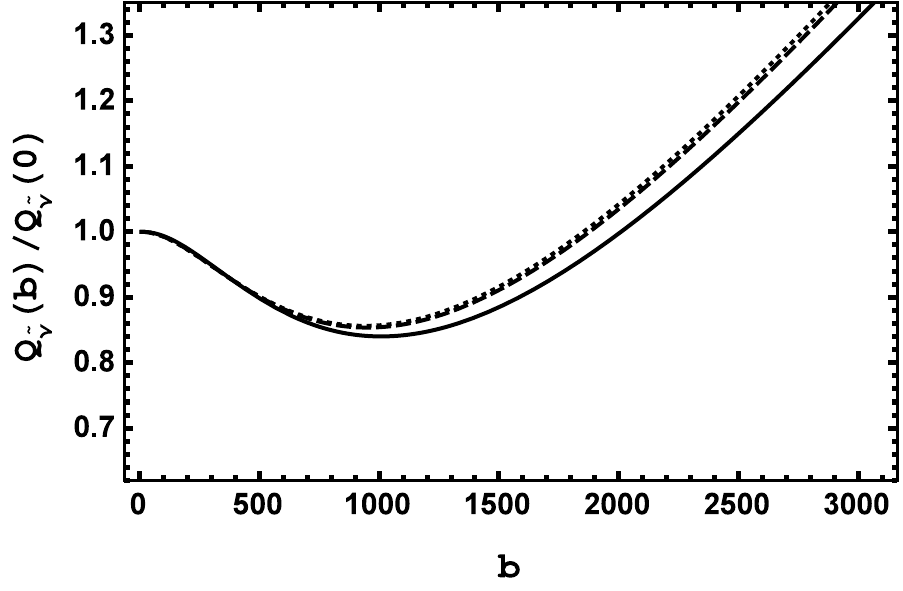} 
}
\end{minipage}
\caption{
The relative energy emitted in neutrinos (the left panel) and 
antineutrinos (the right panel), 
corresponding to the scaled temperature $t \!=\! 8$ and the electron 
degeneracy parameters $a \!=\! 0$ (solid curves), $a \!=\! 2$ (dashed 
curves), and $a \!=\! 6$ (dotted curve) as functions of the magnetic 
field strength $b \!=\! B / B_0$.
} 
\label{fig:Qmu} 
\end{figure}
Numerical analysis has shown~\cite{Ognev:2016wlq} that such dependences 
for the reactions with neutrino and antineutrino differ substantially 
as shown in Fig.~\ref{fig:Qmu} for the relative energy emitted,
$Q_{\nu, \bar\nu} (b) / Q_{\nu, \bar\nu} (0)$. 
It can be seen from Fig.~\ref{fig:Qmu} that the maximum suppression 
of the reaction with the antineutrino participation is observed for 
the nondegenerate plasma ($a = 0$). However, the dependence of the 
relative energy emitted in these processes on the electron degeneracy 
parameter~$a$ at a fixed temperature is very weak. The relative rate 
of the antineutrino generation demonstrates analogous behavior. 
It should be noted that this holds only for the relative quantities 
while the absolute values considerably depend on this parameter 
in terms of $\varGamma_{\bar\nu}(0)$ and $Q_{\bar\nu}(0)$. 
For the processes with neutrino, the situation is radically different 
because even relative values of the quantities strongly depend on the 
electron degeneracy parameter~$a$. As seen in Fig.~\ref{fig:Qmu}, the 
increase in this parameter leads to a displacement of the minima in the 
relative energy emitted in neutrino towards larger values of the field, 
and the minimum itself becomes deeper. The relative rate of the neutrino 
production behaves analogously but the depth of the minimum of this quantity 
is slightly smaller. Thus, the magnetic field suppresses the beta-processes 
with the neutrino participation the most strongly, and this suppression 
is the highest in a degenerate medium.

\begin{figure}[htbp]
\vspace{2mm}
\begin{minipage}[t]{0.49\textwidth}
\centering{
\includegraphics[width=0.95\textwidth]
{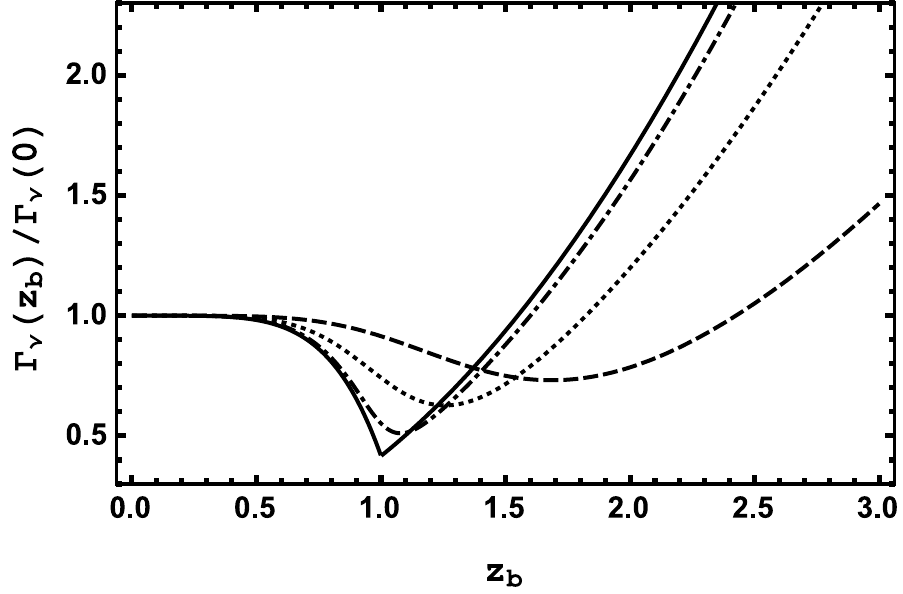} 
}
\end{minipage}
\hfill
\begin{minipage}[t]{0.49\textwidth}
\centering{
\includegraphics[width=0.95\textwidth]
{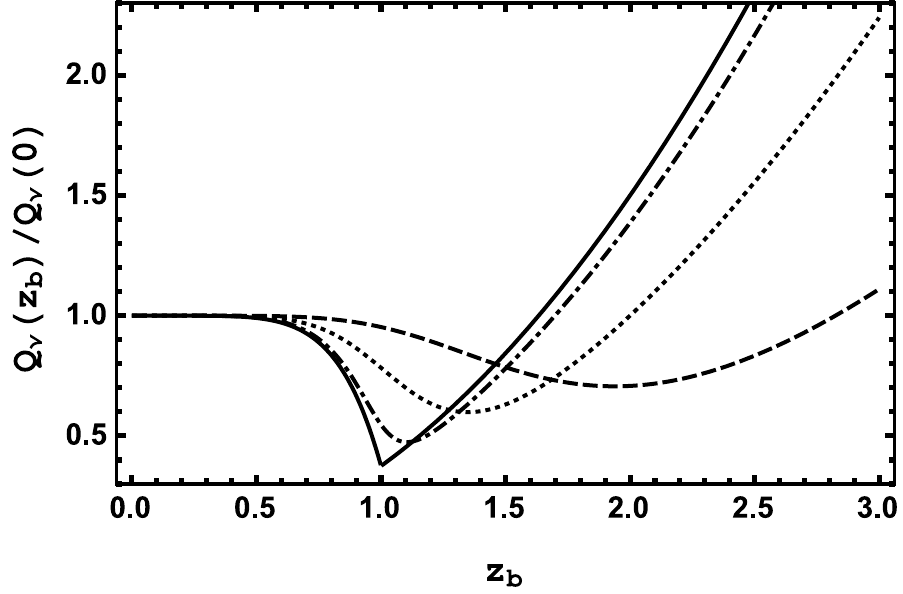} 
}
\end{minipage}
\caption{
The relative rate of the neutrino production (the left panel) 
and energy emitted in this process (the right panel), corresponding 
to the scaled temperature $t \!=\! 8$ and 
degeneracy parameters $a \!=\! 4$ (dashed curves), $a \!=\! 8$ (dotted 
curves), and $a \!=\! 20$ (dash-dotted curves) as functions of the 
dimensionless parameter $z_b \!=\! \sqrt{2eB} / \mu_e$. 
Solid curves correspond to the asymptotic form of the degenerate plasma.
} 
\label{fig:GQnz} 
\end{figure}

In the case of the strong degeneracy of the electron-positron plasma, 
the following simple analytic expressions can be obtained for the 
integrals $J_2 (y_b, a_e)$ and $J_3 (y_b, a_e)$~\cite{Ognev:2016wlq}:
\begin{equation}
\label{eq:J2-J3-approx}
\begin{gathered}
J_2 (y_b, a_e) 
\approx 4/5 \, a_e^5 - 7/15 \, y_b^5 , 
\quad 
J_3 (y_b, a_e) 
\approx 2/3 \, a_e^6 - 5/12 \, y_b^6 , 
\quad  y_b \leqslant a_e ,
\\
J_2 (y_b, a_e) 
\approx 1/3 \, y_b^2 \, a_e^3 , 
\quad 
J_3 (y_b, a_e) 
\approx 1/4 \, y_b^2 \, a_e^4  , 
\quad  y_b \geqslant a_e .
\end{gathered}
\end{equation}
It can easily be seen that the relative neutrino production rate, 
$\varGamma_\nu(z_b) / \varGamma_\nu(0)$, and energy emitted, 
$Q_\nu(z_b) / Q_\nu(0)$, depend only on one variable 
$z_b \!\!=\!\! y_b / a_e \!\!=\!\! \sqrt{2b} \, m_e / \mu_e$, 
and this dependence for the degenerate plasma is universal 
as regards~$\mu_e$. The transition to such an asymptotic 
behavior is demonstrated in Fig.~\ref{fig:GQnz}.
The expressions obtained for $J_n (y_b, a_e)$ in the strongly degenerate 
plasma also show that the largest suppression of the beta-processes with 
neutrino by the magnetic field occurs for the strength
$B_\nu \!\approx\! \mu_e^2 / (2 m_e^2) \, B_0 $, 
and the magnitude of the suppression relative to the field-free value 
is $\varGamma_\nu (B_\nu) / \varGamma_\nu (0) = 5/12 \approx 0.42$
for the neutrino-production rate and 
$Q_\nu (B_\nu) / Q_\nu (0) = 3/8 \approx 0.38$ 
for the energy emitted.

\begin{figure}[htbp]
\vspace{2mm}
\begin{minipage}[t]{0.49\textwidth}
\centering{
\includegraphics[width=0.95\textwidth]
{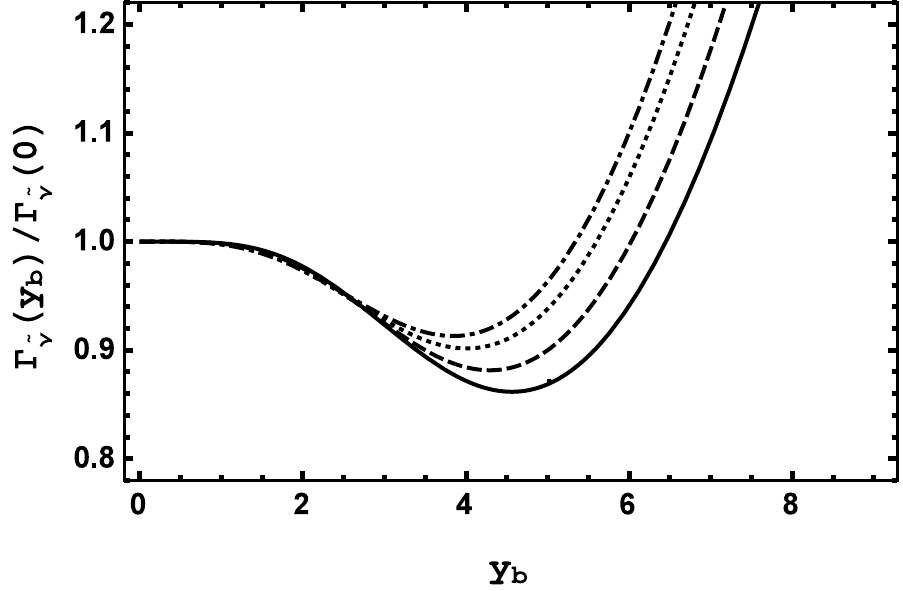} 
}
\end{minipage}
\hfill
\begin{minipage}[t]{0.49\textwidth}
\centering{
\includegraphics[width=0.95\textwidth]
{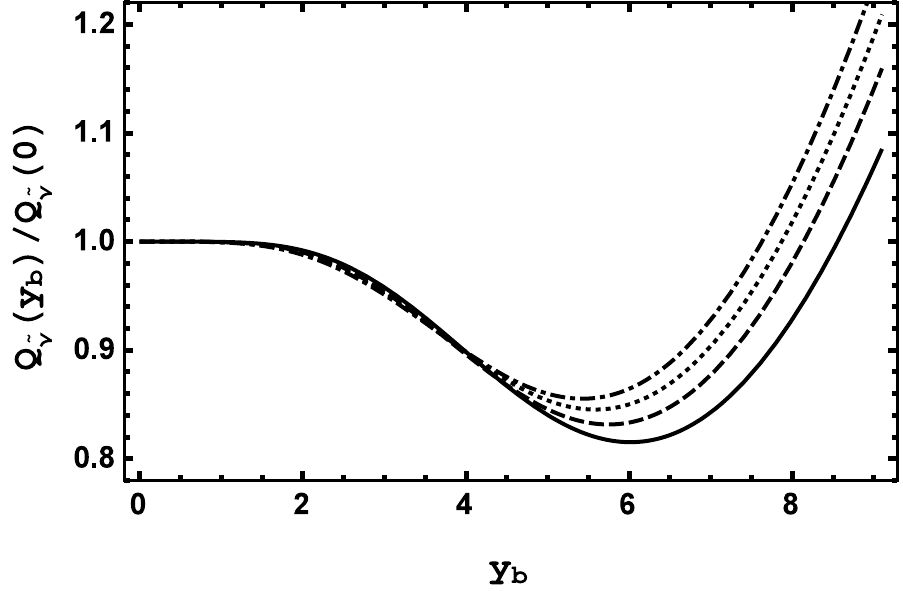} % ??? ??????? ????? ??? ??????????
}
\end{minipage}
\caption{
The relative rate of the antineutrino production (the left panel) 
and energy emitted in this process (the right panel) as functions 
of the dimensionless parameter $y_b \!=\! \sqrt{2b} / t$. Solid 
curves correspond to the asymptotic form of an ultrarelativistic 
nondegenerate plasma ($a \!=\! 0$) in which process~(\ref{proc:ne}) 
has the most strong suppression by the magnetic field. The remaining 
curves correspond to the following dimensionless parameters: 
$t \!=\! 12$ and $a \!=\! 0$ (dashed curves), $t \!=\! 12$ and 
$a \!=\! 3$ (dotted curves), and $t \!=\! 8$ and $a \!=\! 3$ 
(dash-dotted curves).
} 
\label{fig:GQanz} 
\end{figure}

For the beta-processes with the antineutrino participation, the convenient 
variable is the dimensionless parameter $y_b \!=\! \sqrt{2b} / t$.
Numerical calculations show~\cite{Ognev:2016wlq} that the relative rate 
of the antineutrino production, 
$\varGamma_{\bar\nu} (y_b) / \varGamma_{\bar\nu} (0)$,   
and energy, $Q_{\bar\nu} (y_b) / Q_{\bar\nu} (0)$, 
emitted in such processes as functions of this variable weakly depend 
on the temperature of the medium and the plasma degeneracy parameter 
as shown in Fig.~(\ref{fig:GQanz}). 
The same figure depicts the asymptotic form of the nondegenerate 
($a_e \!=\! 0$) ultrarelativistic plasma in which, as noted above, 
the suppression this process by the magnetic field is the strongest. 
It follows from the results of numerical calculations~\cite{Ognev:2016wlq} 
that the maximum suppression of the rate of the antineutrino 
production occurs in the magnetic field with the strength 
$B_{\varGamma_{\bar\nu}} \!\approx\! 10 \, t^2 \, B_0$, 
and the magnitude of the suppression in this case is 
$\varGamma_{\bar\nu} (B_{\varGamma_{\bar\nu}}) / \varGamma_{\bar\nu} (0) \simeq 0.86$. 
The maximum suppression of the energy emitted in antineutrinos 
is shifted towards stronger fields,  
$B_{Q_{\bar\nu}} \!\approx\! 18 \, t^2 \, B_0$, 
and the suppression itself is slightly stronger, 
$Q_{\bar\nu} (B_{Q_{\bar\nu}}) / Q_{\bar\nu} (0) \simeq 0.82$.
Comparison with analogous quantities obtained for processes with the 
neutrino participation shows that the maximum suppression of these 
reactions with the neutrino participation by the magnetic field 
occurs in the nondegenerate medium, and the suppression itself 
is approximately half as strong.

\section*{Conclusions}
\label{concl}

In this paper, we have analyzed the beta-processes~(\ref{proc:pe})--(\ref{proc:n}) 
in the medium of supernovae and accretion disks with an arbitrary magnitude 
of the magnetic field. The results obtained for the transparent medium 
were used to calculate the neutrino and antineutrino production rates, the energy 
emitted in these processes, and the macroscopic momentum transferred to the medium. 

The simplest dependence on the magnetic field was obtained for the macroscopic 
momentum transferred to the matter. It was shown that in the case of an transparent 
medium, it increases linearly with the magnetic field and is directed along 
the field-strength lines. It was demonstrated that this effect is most significant 
in the degenerate electron-positron plasma. Unlike the momentum transferred, the 
rates of the beta-processes and energy emitted in them exhibit a more complex 
dependence on the magnetic field. In the limit of the strong magnetic field, these 
quantities, as well as the momentum transferred, increase linearly with the growth 
of the magnetic field. It was shown, however, that for the values $B \lesssim 10^{15}$~G 
which are typical for supernovae and accretion disks, the values of these quantities 
are suppressed as compared to the field-free case. The maximum suppression of the 
reactions with the neutrino participation occurs in the degenerate medium in the 
field with the strength $B_\nu \sim \mu_e^2 / (2 m_e^2) \, B_0$ which falls into 
the range of values typical of the objects studied. 
%  the case of moderate degeneracy of the plasma. ???? 
The suppression can reach~2.5 times as compared to the field-free case. 
The effect of the magnetic field on the antineutrino processes is weaker and its 
maximum suppression is~1.2 times as compared to the field-free case. However, such 
a suppression occurs in the nondegenerate plasma (i.\,e., in regions of supernovae 
and accretion disks with a moderate density). The magnetic field corresponding 
to the maximum suppression of these processes is $B_{\bar\nu} \sim 10 \, t^2 \, B_0$
which also falls in the range of values typical of the objects considered at moderate 
temperatures. Thus, the suppression of the processes with the neutrino and antineutrino 
participation, by the magnetic field occurs in different parts of supernovae and 
accretion disks, and the effect itself can be significant.

It should be noted that the form of the dependence of the rates of the beta-processes 
and the energy-momentum transferred in these processes on the magnetic field and 
parameters of the matter presented here does not change significantly in the more 
general case of a medium partly transparent to neutrinos and antineutrinos.

I'd like thank to Thomas Janka for numerous discussions on this topic 
and possible extensions of this research.   
It is a pleasure to thank G.\,S.~Bisnovatyi-Kogan, Alexander Parkhomenko and Alexandra Dobrynina 
for stimulating discussions and critical comments. 
This study was partially supported 
by the Russian Foundation for Basic Research (project no. 15-02-06033-a)
and by the Michael Lomonosov Program (project no. 1.721.2016/2.2).


\begin{thebibliography}{10}
\def\selectlanguageifdefined#1{
\expandafter\ifx\csname date#1\endcsname\relax
\else\selectlanguage{#1}\fi}
\providecommand*{\href}[2]{{\small #2}}
\providecommand*{\url}[1]{{\small #1}}
\providecommand*{\BibUrl}[1]{\url{#1}}
\providecommand{\BibAnnote}[1]{}
\providecommand*{\BibEmph}[1]{#1}
\ProvideTextCommandDefault{\cyrdash}{\iflanguage{russian}{\hbox
  to.8em{--\hss--}}{\textemdash}}
\providecommand*{\BibDash}{\ifdim\lastskip>0pt\unskip\nobreak\hskip.2em plus
  0.1em\fi
\cyrdash\hskip.2em plus 0.1em\ignorespaces}
\renewcommand{\newblock}{\ignorespaces}

\bibitem{Kumar:2014upa}
\selectlanguageifdefined{english}
\BibEmph{Kumar~P., Zhang~B.} {The physics of gamma-ray bursts~\& relativistic
  jets}~//
  \href{http://dx.doi.org/10.1016/j.physrep.2014.09.008}{\BibEmph{Phys. Rept.}}
  \BibDash
\newblock 2014. \BibDash
\newblock Vol. 561. \BibDash
\newblock P.~1--109. \BibDash
\newblock 1410.0679.

\bibitem{MacFadyen:1998vz}
\selectlanguageifdefined{english}
\BibEmph{MacFadyen~A.~I., Woosley~S.~E.} {Collapsars: Gamma-Ray Bursts and
  Explosions in ``Failed Supernovae''}~//
  \href{http://dx.doi.org/10.1086/307790}{\BibEmph{Astrophys. J.}} \BibDash
\newblock 1999. \BibDash
\newblock Vol. 524. \BibDash
\newblock P.~262--289. \BibDash
\newblock astro-ph/9810274.

\bibitem{Rosswog:2015nja}
\selectlanguageifdefined{english}
\BibEmph{Rosswog~S.} {The multi-messenger picture of compact binary mergers}~//
  \href{http://dx.doi.org/10.1142/S0218271815300128}{\BibEmph{Int.~J. Mod.
  Phys.}} \BibDash
\newblock 2015. \BibDash
\newblock Vol. D24, no.~5. \BibDash
\newblock P.~1530012. \BibDash
\newblock 1501.02081.

\bibitem{Barkov:2009zq}
\selectlanguageifdefined{english}
\BibEmph{Barkov~M.~V., Komissarov~S.~S.} {Close Binary Progenitors of Long
  Gamma Ray Bursts}~//
  \href{http://dx.doi.org/10.1111/j.1365-2966.2009.15792.x}{\BibEmph{Mon. Not.
  Roy. Astron. Soc.}} \BibDash
\newblock 2010. \BibDash
\newblock Vol. 401. \BibDash
\newblock P.~1644--1656. \BibDash
\newblock 0908.0695.

\bibitem{Bruenn:1985en}
\selectlanguageifdefined{english}
\BibEmph{Bruenn~S.~W.} {Stellar core collapse: Numerical model and infall
  epoch}~// \href{http://dx.doi.org/10.1086/191056}{\BibEmph{Astrophys. J.
  Suppl.}} \BibDash
\newblock 1985. \BibDash
\newblock Vol.~58. \BibDash
\newblock P.~771--841.

\bibitem{Moiseenko:2015}
\selectlanguageifdefined{english}
\BibEmph{Moiseenko~S.~G., Bisnovatyi-Kogan~G.~S.} {Development of the
  magneto-differential-rotational instability in magnetorotational
  supernova}~//
  \href{http://dx.doi.org/10.1134/S1063772915070069}{\BibEmph{Astron. Rep.}}
  \BibDash
\newblock 2015. \BibDash
\newblock Vol.~59, no.~4. \BibDash
\newblock P.~573--580.

\bibitem{Sawai:2015tsa}
\selectlanguageifdefined{english}
\BibEmph{Sawai~H., Yamada~S.} {The Evolution and Impacts of Magnetorotational
  Instability in Magnetized Core-Collapse Supernovae}~//
  \href{http://dx.doi.org/10.3847/0004-637X/817/2/153}{\BibEmph{Astrophys. J.}}
  \BibDash
\newblock 2016. \BibDash
\newblock Vol. 817. \BibDash
\newblock P.~153. \BibDash
\newblock 1504.03035.

\bibitem{Zalamea:2010ax}
\selectlanguageifdefined{english}
\BibEmph{Zalamea~I., Beloborodov~A.~M.} {Neutrino Heating Near Hyper-Accreting
  Black Holes}~//
  \href{http://dx.doi.org/10.1111/j.1365-2966.2010.17600.x}{\BibEmph{Mon. Not.
  Roy. Astron. Soc.}} \BibDash
\newblock 2011. \BibDash
\newblock Vol. 410. \BibDash
\newblock P.~2302--2308. \BibDash
\newblock 1003.0710.

\bibitem{Gvozdev:2002ta}
\selectlanguageifdefined{english}
\BibEmph{Gvozdev~A.~A., Ognev~I.~S.} {Efficiency of electron positron pair
  production by neutrino flux from accretion disk of a Kerr black hole}~//
  \href{http://dx.doi.org/10.1134/1.1421403}{\BibEmph{JETP Lett.}} \BibDash
\newblock 2001. \BibDash
\newblock Vol.~74. \BibDash
\newblock P.~298--301. \BibDash
\newblock [Pisma Zh. Eksp. Teor. Fiz. 74, 330 (2001)]. astro-ph/0201346.

\bibitem{Gvozdev:2002nu}
\selectlanguageifdefined{english}
\BibEmph{Gvozdev~A.~A., Ognev~I.~S.} {Neutrino interaction with nucleons in the
  envelope of a collapsing star with a strong magnetic field}~//
  \href{http://dx.doi.org/10.1134/1.1493155}{\BibEmph{J.~Exp. Theor. Phys.}}
  \BibDash
\newblock 2002. \BibDash
\newblock Vol.~94. \BibDash
\newblock P.~1043--1056. \BibDash
\newblock [Zh. Eksp. Teor. Fiz. 121, 1219 (2002)]. astro-ph/0403011.

\bibitem{Gvozdev:2005hz}
\selectlanguageifdefined{english}
\BibEmph{Gvozdev~A.~A., Ognev~I.~S.} {Influence of a strong magnetic field on
  the neutrino heating of a supernova shock}~//
  \href{http://dx.doi.org/10.1134/1.1958108}{\BibEmph{Astron. Lett.}} \BibDash
\newblock 2005. \BibDash
\newblock Vol.~31. \BibDash
\newblock P.~442--445.

\bibitem{Ognev:2016wlq}
\selectlanguageifdefined{english}
\BibEmph{Ognev~I.~S.} {Effect of magnetic field on beta processes in a
  relativistic moderately degenerate plasma}~//
  \href{http://dx.doi.org/10.1134/S106377611610006X}{\BibEmph{J. Exp. Theor.
  Phys.}} \BibDash
\newblock 2016. \BibDash
\newblock Vol. 123, no.~4. \BibDash
\newblock P.~643--665. \BibDash
\newblock [Zh. Eksp. Teor. Fiz. 150, 744 (2016)].

\bibitem{Olive:2016xmw}
\selectlanguageifdefined{english}
\BibEmph{Patrignani~C. et~al.} {Review of Particle Physics}~//
  \href{http://dx.doi.org/10.1088/1674-1137/40/10/100001}{\BibEmph{Chin.
  Phys.}} \BibDash
\newblock 2016. \BibDash
\newblock Vol. C40, no.~10. \BibDash
\newblock P.~100001.

\end{thebibliography}
\end{document}